\documentclass[8.5pt,twoside,twocolumn]{article}
\usepackage[super,sort&compress,comma]{natbib} 
\usepackage[version=3]{mhchem}
\usepackage{balance}
\usepackage{times,mathptm}
\usepackage{sectsty}
\usepackage{graphicx} 
\usepackage{lastpage}
\usepackage[format=plain,justification=raggedright,singlelinecheck=false,font=small,labelfont=bf,labelsep=space]{caption} 
\usepackage{fancyhdr}
\begin{document}

\thispagestyle{plain}
\fancypagestyle{plain}{
\renewcommand{\headrulewidth}{1pt}}
\renewcommand{\thefootnote}{\fnsymbol{footnote}}
\renewcommand\footnoterule{\vspace*{1pt}%
\hrule width 3.4in height 0.4pt \vspace*{5pt}} 
\setcounter{secnumdepth}{5}

\makeatletter 
\renewcommand\@biblabel[1]{#1}            
\renewcommand\@makefntext[1]%
{\noindent\makebox[0pt][r]{\@thefnmark\,}#1}
\makeatother 
\renewcommand{\figurename}{\small{Fig.}~}
\sectionfont{\large}
\subsectionfont{\normalsize} 

\fancyfoot{}
\fancyfoot[RO]{\footnotesize{\sffamily{1--\pageref{LastPage} ~\textbar  \hspace{2pt}\thepage}}}
\fancyfoot[LE]{\footnotesize{\sffamily{\thepage~\textbar\hspace{3.45cm} 1--\pageref{LastPage}}}}
\fancyhead{}
\renewcommand{\headrulewidth}{1pt} 
\renewcommand{\footrulewidth}{1pt}
\setlength{\arrayrulewidth}{1pt}
\setlength{\columnsep}{6.5mm}
\setlength\bibsep{1pt}

\twocolumn[
  \begin{@twocolumnfalse}
\noindent\LARGE{\textbf{Kinetic dielectric decrement revisited: phenomenology of finite ion concentrations$^\dag$}}
\vspace{0.6cm}

\noindent\large{\textbf{Marcello Sega,$^{\ast}$\textit{$^{a}$} Sofia Kantorovich,\textit{$^{b\ddag}$} and
Axel Arnold\textit{$^{c}$}}}\vspace{0.5cm}


 \end{@twocolumnfalse} \vspace{0.6cm}

  ]

\noindent\textbf{
With the help of a recently developed non-equilibrium approach, we
investigate the ionic strength dependence of the Hubbard--Onsager
dielectric decrement. We compute the depolarization of water molecules
caused by the motion of ions in sodium chloride solutions from the
dilute regime (0.035 M) up close to the saturation concentration (4.24 M),
and find that the kinetic decrement displays a strong nonmonotonic
behavior, in contrast to the prediction of available models. We
introduce a phenomenological modification of the Hubbard--Onsager continuum theory,
that takes into account the screening due to the ionic cloud at mean
field level, and, is able to describe the 
kinetic decrement at high concentrations including the presence of a pronounced minimum.
}
\section*{}
\vspace{-1cm}


\footnotetext{\textit{$^{a}$~University of Vienna, Department of Computational Biological Chemistry, W\"ahringer Strasse 17, 1090 Vienna, Austria ; E-mail: marcello.sega@univie.ac.at}}
\footnotetext{\textit{$^{b}$~University of Vienna, Faculty of Physics, Boltzmanngasse 5, 1090 Vienna, Austria}}
\footnotetext{\textit{$^{c}$~Institute for Computational Physics, Universit\"at Stuttgart,Allmandring 3, 70569 Stuttgart, Germany}}


More than thirty years ago, in what was one of the last articles
written by L. Onsager, he and J. P. Hubbard made a captivating
prediction that has eluded direct observation until now. They stated that in a
saline solution, due to the motion of ions, polar solvent molecules
should show a tendency to orient against any external, static
electric field, in apparent contradiction with
electrostatics.\cite{hubbard77a,hubbard78a,hubbard79a}  According to the continuum model
of Hubbard and Onsager, the rotational current induced in the solvent
by ionic currents should generate a net solvent depolarization that
survives in the zero frequency limit. As a consequence, a decrement of
the \emph{static} permittivity of the solution should be observed,
even though the effect is purely \emph{dynamic}, and as such can not be
explained in terms of molecular configurations only.  To date,
however, no direct experimental proof of the kinetic decrement
exists, because its detection is complicated by the presence of
dielectric saturation, from which it can not be easily separated.\cite{winsor1982dielectric,wei92,nortemann1997dielectric,kaatze97,kaatze2011bound}
A quantitative picture of the kinetic contribution to the dielectric
decrement is therefore key to the investigations of ion solvation
properties, which rely on a correct estimate of the static contribution
of the
decrement.\cite{tielrooij2010cooperativity,buchner1999dielectric,nandi2000dielectric}

The continuum theory of the kinetic decrement predicts that the
static permittivity $\epsilon_0$ of a solvent should change upon
addition of salt by an amount
\begin{equation}\Delta\epsilon_{HO}=-4 \pi
p \sigma \tau (\epsilon_0-\epsilon_\infty)/\epsilon_0, \label{eq:HO}
\end{equation}
due to a subtle interplay between ion motion and rotational orientation of the solvent molecules. Here, $\sigma$ denotes the conductivity of the solution
and $\tau$ is the time constant of the Debye relaxation process
that characterizes the dielectric susceptibility of the solvent,
and $\epsilon_\infty$ is the infinite frequency dielectric constant.
\cite{hubbard77a,hubbard78a} The factor $p$ can take
values between 2/3 and 1, depending on the type of boundary condition
at the surface of the ion (full slip and no slip, respectively).
Strictly speaking, the continuum theory is valid only in the infinite dilution limit, and for large
ionic radii.\cite{hubbard79a}

Despite these limitations, the formula
for the decrement bears an enthralling elegance, and 
explains qualitatively the dependence of the dielectric permittivity
of electrolyte solutions on their conductivity, even well within
the concentrate solution regime.\cite{hubbard77pnas} However, the kinetic
decrement is not the only effect  that is expected to lower the
dielectric permittivity of electrolyte solutions. The strong local
electric field in the vicinity of the ions tends to polarize solvent
molecules more than any external electric field in the linear regime.
Such a high field  saturates the dielectric response of solvent
molecules next to ions, effectively reducing the dielectric
permittivity of the solution.  This effect depends on the salt
concentration $c$ and, implicitly, on the conductivity $\sigma$.
For this reason it becomes hard, if not impossible, to separate
the kinetic contribution from dielectric saturation
experimentally.\cite{winsor1982dielectric,wei92,kaatze97,kaatze2011bound}
This situation prevents not only a direct observation of the kinetic
decrement, but also a precise evaluation of the effect of
saturation.\cite{nortemann1997dielectric}

Here, we use a non-equilibrium molecular dynamics approach to compute
the kinetic decrement over an unprecedented wide range of concentrations,
which is not accessible with conventional, equilibrium approaches.\cite{chandra00x}
Moreover, we present a simple phenomenological theory that gives a quantitative account of
the features of the kinetic decrement at higher concentrations.

\begin{figure}
\includegraphics[width=\columnwidth]{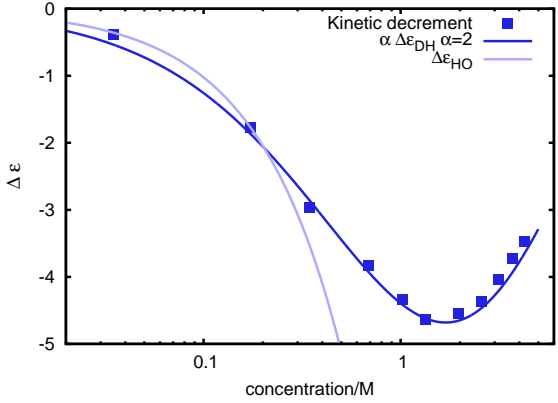}
\caption{Kinetic contribution to the static dielectric permittivity. Squares: simulation result; light line: Hubbard--Onsager theory, $\Delta\epsilon_{HO}(c)$; dark line:  $\alpha\,\Delta\epsilon_{DH}(c)$, with $\alpha=2$}
\label{fig:decrement}
\end{figure}

The kinetic decrement can be seen as the sum of two complementary
contributions: the first is the depolarization due to
the moving ions, which exerts a torque on the solvent molecules; the
second, more subtle effect, is a change in the imaginary part of the ion
conductivity, a lag in the response of ions induced by the rotation of
solvent molecules which are orienting along the external electric field.
The two contributions must have exactly the same value, as a consequence
of Onsager reciprocal relations. A straightforward way to see this is
through the Green--Kubo expression for the kinetic decrement.
The change in solvent polarization current $\mathbf{J}_p$ due to the
ionic one $\mathbf{J}_i$ leads to a contribution to the conductivity
spectrum $\Delta\sigma_{pi}(\omega)=\beta/(3V) \int_0^\infty \exp(i\omega
t)\left\langle\mathbf{J}_p(t)\mathbf{J}_i(0)\right\rangle \mathrm{d}t$,
 where
$\beta$ is the inverse thermal energy, $V$ the simuation box volume, and $\left\langle\cdot\right\rangle$ is a suitable ensemble average.
This change in conductivity reflects a change in permittivity, since 
$\epsilon(\omega)-1
= i 4\pi\sigma(\omega)/\omega$,\cite{hansen86a} and results in the first contribution to the kinetic decrement 
 $\Delta\epsilon_{pi} = \lim_{\omega\to0}4\pi
 i\Delta\sigma_{pi}(\omega)/\omega$. Owing to the symmetry of the
 current cross-correlation function, the second contribution , which
 originates from the action of the rotating solvent molecules on
the ions, is $\Delta\epsilon_{ip}=\Delta\epsilon_{pi}$. The  total
kinetic decrement is therefore twice the first contribution, $\Delta\epsilon=2\Delta\epsilon_{pi}$.

However elegant, the Green--Kubo expression is not very much suited  for
the computation of the kinetic decrement, because the signal-to-noise
 ratio at extreme dilutions would be too small for any practical
purposes. A much more efficient way to compute $\Delta\epsilon_{pi}$
consists instead in applying an external fictitious field $\mathbf{E}_f$,
that couples to the ions only, and in measuring the resulting
polarization of the solvent $\mathbf{P}$, so that $\Delta\epsilon_{pi}=4\pi
P/( V E_f )$\cite{sega14a}. This is evidently the out-of-equilibrium counterpart of
the Green-Kubo formula for $\Delta\epsilon_{pi}$, because $\mathbf{J}_i$ is
the current that couples to the external field $\mathbf{E}_f$, and $\mathbf{J}_p$
the one generating the polarization $\mathbf{P}$. Even rather intense
fictitious fields do not drive the system out of the linear regime, and
allow to collect meaningful statistics also for very dilute solutions with
relatively short simulation runs, making this  non-equilibrium approach
the key to calculating the kinetic decrement over an unprecedented wide
range of concentrations.  We applied this non-equilibrium calculation
to an aqueous solution of sodium chloride at 11 different salt
concentrations. In our simulations we model water molecules using
the three-sites SPC/E potential\cite{berendsen87a} and sodium and
chloride ions using the thermodynamics consistent Kirkwood--Buff
potential.\cite{weerasinghe03a} The salt concentration $c$ varies from
0.035 to 4.24 M, keeping the water content fixed at 1621 molecules
per simulation box and changing the number of salt pairs from 1 to
140. We kept constant temperature (300 K) and pressure (1 atm) using
the Nos\'e--Hoover\cite{nose84a,hoover85a} and
Parrinello--Rahman\cite{parrinello81a} algorithms with relaxation
times of 5 ps.  Electrostatic interactions were computed using the
smooth particle mesh Ewald method\cite{essmann95a} with tin-foil
boundary conditions, a 4-th order interpolation spline on a grid
with spacing not larger than 0.12 nm and a relative interaction
strength of $10^{-5}$ at 0.9 nm. We switched the short range part
of the electrostatic interaction and the Lennard-Jones smoothly to
zero between 0.9 to 1.2 nm using a fourth-degree polynomial.
Simulations were performed with an in-house modified version of
gromacs\cite{gromacs4} for the on-line calculation of the currents
associated to the different species, and used an integration time
step of 1 fs.

\begin{figure}
\includegraphics[width=\columnwidth]{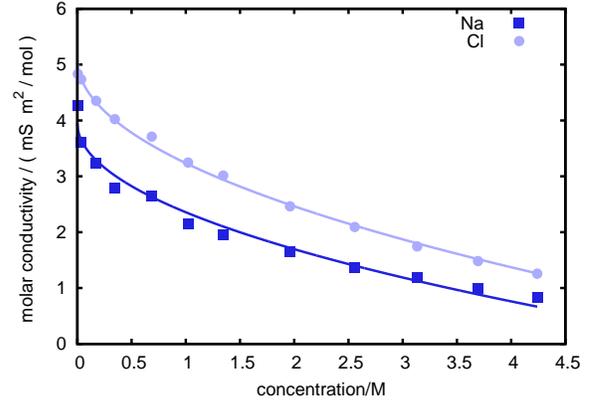}
\caption{Molar conductivity of sodium and chloride ions, as a function of the concentration. Squares: sodium; Circles: chloride; Solid lines: best fit to Kohlrausch law.}
\label{fig:conductivity}
\end{figure}

In Figure~\ref{fig:decrement} we show the kinetic decrement as a function of the salt
concentration, as measured in our simulations. The kinetic decrement
shows a marked non monotonous dependence on the concentration,
with a clear minimum right before $c=2$~M.  To test the Hubbard--Onsager
formula, Eq.(\ref{eq:HO}), we calculated also the molar conductivity
$\lambda$ of the solution at the different concentrations, as reported in
Figure~\ref{fig:conductivity}. A best fit to the Kohlraush law,
$\lambda(c)=\lambda_0-K\sqrt{c}$, allows us to extrapolate the molar
conductivity to infinite dilution, and estimate the limiting
molar conductivity $\lambda_0$ separately for the sodium and chloride
ions. Therefore, we can write the Hubbard--Onsager decrement for
the mixture of sodium and chloride in the form $\Delta\epsilon_{HO} =
- 4\pi p  \tau (\lambda^{Na}_0 + \lambda^{Cl}_0 ) c
(\epsilon_0-\epsilon_\infty)/\epsilon_0$. The Hubbard--Onsager decrement so calculated (Fig.\ref{fig:decrement}, light line)
is not compatible with the simulation data above a concentration
of 0.2 M. The presence of a pronounced minimum and the subsequent
increase of $\Delta\epsilon$ can not be explained even qualitatively
with the continuum Hubbard--Onsager theory.

Physical intuition suggests that the local field of the ions, which
determines the torque on water molecules, should be screened by the
presence of oppositely charged ions in its vicinity. To formalize
this, we introduce a mean-field correction to the Hubbard--Onsager
theory, along the lines of the Debye--H\"uckel theory. The crucial
step in the Hubbard--Onsager theory is the calculation of the
rotational current $\mathbf{J}_R$ induced in the polar medium by
an ion travelling with a speed $\mathbf{u}$.  The coupling between
electrostatics and Navier--Stokes equations allows to express the
rotational current as a functional of the local field generated by
the ion, $\mathbf{E}_0$, as $\mathbf{J}_R  = \int (\chi/2\epsilon_0)
\mathbf{E}_0 \times \left( \nabla\times\mathbf{v} \right) dV$, where
$\mathbf{v}$ is the velocity field of the solvent surrounding the
ion and $\chi$ its dielectric susceptibility.\cite{hubbard77a} For
large ionic radii $R$, the velocity field can be approximated by
the Stokes solution, $\mathbf{v}(\mathbf{r}) =  (3R/4r^3) \left[
r^2 \mathbf{u} + (\mathbf{r}\cdot\mathbf{u})\mathbf{r} \right]$.
If, instead of using the Coulomb field, we use the Debye--H\"uckel
one, $\mathbf{E}_0 = (q/\epsilon_0 r^3)\exp({-\kappa r})(1+\kappa
r) \mathbf{r}$, the rotational current can be evaluated analytically
as $\mathbf{J}_R = (2\pi/\epsilon_0) u \chi q \exp({-\kappa R })$.
Here $\kappa=\sqrt{\beta c  e^2 /(2 \pi \epsilon_0)}$ is the inverse
Debye screening length.

Since the ratio between the ion speed $u$ and the driving electric field $E_x$ is $u/E_x = \sigma/q$, it is possible express the dielectric decrement (which we denote here as $\Delta\epsilon_{DH}$, the suffix standing for Debye--H\"uckel) in terms of the rotational current
\begin{equation}
\Delta\epsilon_{DH} = \lim_{\omega\to 0} 4\pi \frac{\sigma^{''}}{\omega} = \lim_{\omega\to 0} \frac{4 \pi }{\omega} J^{''}_R/E_x.
\end{equation}
Here, the imaginary part of a quantity is denoted by double-primes.
The susceptibility of the dipolar medium is assumed to be characterized by a single Debye relaxation, so that $4\pi \chi(\omega) = (\epsilon_0-\epsilon_\infty)(1+i\omega\tau_D)$, from which one derives
\begin{equation}
\Delta\epsilon_{DH} =  4\pi \sigma\tau \frac{\epsilon_0-\epsilon_\infty}{\epsilon_0} e^{-\kappa R }p.\label{eq:final}
\end{equation}
The solution resembles the classical Hubbard--Onsager one, but
features an additional factor $\exp({-\kappa R })$ which depends on the
(effective) ion size. This difference is an important one, because it
shows that even at the mean field level there is an additional length scale, the screening length $\kappa^{-1}$, that governs the non-monotonous behaviour of the kinetic decrement.

In Figure~\ref{fig:decrement} we compare the simulation results
with the mean field result Eq. (\ref{eq:final}),  summed over the
contributions of the two ionic species, and multiplied by a scaling
factor $\alpha$. The parameter $\alpha$ takes into account in a
phenomenological way the effect of ionic correlations arising at
high salt concentration.  Very good agreement at high concentration
has been achieved when $\alpha=2$ (dark curve). As an effective
ionic radius we used the size of the first hydration shell of the
ion, defined as the sum of the  position of the first minimum in
the ion-water radial distribution function and of the Lennard-Jones
diameter of a water molecule.  The relaxation time $\tau$ has been
computed from a fit of the Debye process $\chi(\omega)$ to the
spectrum of the pure solvent, and the solution conductivity has
been calculated from the limiting molar ones as for the Hubbard--Onsager
case. One should notice that the relaxation of water is not described
by a single Debye process, and often a Cole-Cole relaxation or two
Debye processes\cite{barthel1990dielectric} are used to fit
experimental data. However, the dominant contribution at lower
frequencies comes from the main Debye relaxation, which is the one
we are using here to calculate the decrement.

As a final remark, we note that due to the presence of the scaling
factor $\alpha$, the curve does not converge, for  $\kappa R \ll
1$, to the solution of Hubbard and Onsager, the latter being a better
approximation at low concentrations.  Nevertheless, the simulation
data shows that the applicability range of the Hubbard--Onsager
theory is limited to concentration smaller than approximately 0.2~M,
a condition which has been often not fulfilled when searching for
experimental evidences of the kinetic decrement.\cite{hubbard77pnas}
Our simulation results thus resolve the doubts which were cast on
the attribution of the measured dielectric decrement\cite{wei92}
in favor of the hypothesis of a static effect arising from dielectric
saturation,\cite{kaatze97} which is definitely the largest contribution
to the dielectric decrement.

M.S. acknowledges support from the European
Community's Seventh Framework Programme (FP7-PEOPLE-2012-IEF) funded
under grant Nr.~331932 SIDIS.  S.S.K. acknowledges support from
RFBR grants mol-a 1202-31-374 and mol-a-ved 12-02-33106, from the
Ministry of Science and Education of RF 2.609.2011 and, from Austrian
Science Fund (FWF): START-Projekt Y 627-N27.  The authors thank Christian
Schr\"oder and Othmar Steinhauser for useful discussions. A.A. would
like to thank the German Research Foundation (DFG) for financial
support through the Cluster of Excellence in Simulation Technology
(EXC 310/1) at the University of Stuttgart.



\footnotesize{

\providecommand*{\mcitethebibliography}{\thebibliography}
\csname @ifundefined\endcsname{endmcitethebibliography}
{\let\endmcitethebibliography\endthebibliography}{}

}

\end{document}